\documentstyle[12pt]{article}
\begin{document}
{\hskip 12.0cm} AS-ITP-95-31\par
\vspace{1.0ex}
%{\hskip 12.0cm} hep-ph/9509nn\\
\vspace{6ex}
\begin{center}        
{\LARGE \bf $1/m$ corrections to heavy baryon masses
in the heavy quark effective theory sum rules}\\
\vspace{5ex}
{\sc Yuan-ben Dai$^{a}$, Chao-shang Huang$^{a}$,
Chun Liu$^{b, a}$, Cai-dian L\"u$^{b, a}$}\\
\vspace{3ex}
{\it  $^a$ Institute of Theoretical Physics, Academia Sinica}\\
{\it P.O. Box  2735,  Beijing  100080, China
\footnote{Mailing address}}\\
{\it  $^b$ CCAST (World Laboratory) P.O. Box 8730, Beijing, 100080}\\     
\vspace{8.0ex}
{\large \bf Abstract}\\
\vspace{4ex}
\begin{minipage}{130mm}
                                                                       
   The $1/m$ corrections to heavy baryon masses are calculated from the
QCD sum rules within the framework of the heavy quark effective theory.
Numerical results for the heavy baryons are obtained.  The implications
of the results are discussed.\par
\vspace{0.5cm}
\end{minipage}
\end{center}

\newpage
      
   Heavy baryons provide us a testing ground to the Standard Model (SM),
especially to QCD in some aspects.  With the accumulation of the 
experimental data on the heavy baryons, more reliable theoretical 
calculations are needed, although some of them are rather complicated.
Within the framework of the heavy quark effective theory (HQET) which
is a model-independent method, the theoretical analysis to the heavy
baryons containing a single heavy quark is comparatively simple because 
of the heavy quark symmetry [1].  However there are still quantities in 
this framework
which need to be determined from nonperturbative QCD.\par
\vspace{1.0cm}    
   QCD sum rule [2], which is regarded as a nonperturbative method rooted 
in QCD itself, has been used successfully to calculate the properties of
various hadrons.  For instances, besides the light mesons [2], light baryons
were first considered in Ref. [3].  Heavy meson properties
were systematically analyzed
within the HQET [4,5].  Heavy baryons were first discussed in Ref. [6], then
masses and Isgur-Wise function for heavy baryons
were calculated in the HQET to the leading order heavy quark expansion 
in Refs. [7] and [8].  In Ref. [9], the calculation for the heavy baryons 
began 
with the full theory and results of the calculation
were expanded by heavy quark masses.
In this paper, within the framework of the HQET, we study the heavy baryonic
two-point correlators to the subleading order of the heavy quark expansion
by QCD sum rule and obtain results for the heavy baryon
masses to that order.\par
\vspace{1.0cm}    
   In the HQET, the heavy 
quark mass $m_Q$ which is defined perturbatively as the pole mass has been
removed by the field redefinition.  The heavy quark field $h_v$ is defined 
by
\begin{equation}
P_+Q(x)=\exp(-im_Qv\cdot x)h_v(x)~,
\end{equation}
where $P_+=\frac{1}{2}(1+\not \! v)$.  To the order of 
$1/m_Q$, the effective Lagrangian for the heavy quark is [4]
\begin{equation}
{\cal L}_{\rm eff}=\bar{h}_viv\cdot Dh_v+\frac{1}{2m_Q}\bar{h}_v(iD)^2h_v
-\frac{g}{4m_Q}\bar{h}_v\sigma_{\mu\nu}G^{\mu\nu}h_v~.
\end{equation}
As for the $1/m_Q$ terms, the first one still conserves heavy quark spin 
symmetry.  It is the last term which violates the spin symmetry.  The heavy
baryon mass $M$ is expanded as [10]
\begin{equation}
M=m_Q+\bar{\Lambda}+\frac{\delta\Lambda^K}{m_Q}
+\frac{\delta\Lambda^G}{m_Q}<\vec{s}_Q\cdot\vec{j}_l>+O(\frac{1}{m_Q^2})~,
\end{equation}
where $\bar{\Lambda}$ is the heavy baryon mass in the heavy quark limit,
which has been calculated in Ref. [7].  $\delta\Lambda^K$ and
$\delta\Lambda^G$ parameterize the spin-conserved and spin-violated $1/m_Q$
corrections respectively.  All of them characterize the properties of the
light degrees of freedom.  $s_Q$ denotes the heavy quark spin, and $j_l$
stands for the total angular momentum of the light degrees of freedom.
For $\Lambda_Q$ baryon, $\delta\Lambda^G$ term vanishes.  For
$\Sigma_Q^{(*)}$ baryons, both $\delta\Lambda^K$ and $\delta\Lambda^G$ terms
are nonvanishing with $<\vec{s}_Q\cdot\vec{j_l}>=-1$ for $\Sigma_Q$ and
$\frac{1}{2}$ for $\Sigma_Q^*$.\par
\vspace{1.0cm} 
   The heavy baryonic currents $\tilde{j}^v$ have been given in Refs. [6] and
[7] in the rest frame of the heavy baryons.  Generally they can be expressed 
as
\begin{equation}
\tilde{j}^v=\epsilon^{abc}(q_1^{{\rm T}a}C\Gamma\tau q^b_2)\Gamma'h_v^c~,
\end{equation}
where $C$ is the charge conjugate matrix, $\tau$ is a flavor matrix, 
$\Gamma$ and $\Gamma'$ are some gamma matrices, and $a, b, c$ denote the
color indices. $\Gamma$ and $\Gamma'$ can be chosen covariantly as 
\begin{equation}
\Gamma_{\Lambda}=\gamma_5~~~~~~\Gamma'_{\Lambda}=1~,
\end{equation}
for $\Lambda_Q$ baryon;
\begin{equation}
\Gamma_{\Sigma}=\gamma^{\mu}~~~~~~\Gamma'_{\Sigma}=(v^{\mu}
+\gamma^{\mu})\gamma_5~,
\end{equation}
for $\Sigma_Q$ baryon;
\begin{equation}
\Gamma_{\Sigma^*}=\gamma^{\nu}~~~~~~
\Gamma'_{\Sigma^*}=-g^{\mu\nu}+\frac{1}{3}\gamma^{\mu}\gamma^{\nu}
-\frac{1}{3}(\gamma^{\mu}v^{\nu}-\gamma^{\nu}v^{\mu})+\frac{2}{3}
v^{\mu}v^{\nu}~,
\end{equation}
for $\Sigma_Q^*$ baryon.  The choice of $\Gamma$ is not unique.  
Another kind of
baryonic current can be obtained by inserting a factor $\not v$ 
before the $\Gamma$ in Eqs. (5-7).  The
currents given by Eqs. (5-7) are denoted 
as $\tilde{j}^v_1$, and that with $\not v$
insertion as $\tilde{j}^v_2$.
We define the "baryonic decay constant"
$f$ in the HQET as follows
\begin{equation}
\begin{array}{lll}
<0|\tilde{j}^v|\Lambda_Q>&=&f_{\Lambda}u~,\\
% ~~~{\rm for} ~~~{\cal B}=\Lambda_Q, \Sigma_Q~, 
<0|\tilde{j}^v|\Sigma_Q>&=&f_{\Sigma}u~,\\
<0|\tilde{j}^v_{\mu}|\Sigma^*_Q>&=&\frac{1}{\sqrt{3}}f_{\Sigma^*}u_{\mu}~,
\end{array}
\end{equation}
where $u$ is the spinor and  
$u_{\mu}$ is the Rarita-Schwinger spinor in the HQET respectively.  
$f_{\Sigma^*}$ is the same as 
$f_{\Sigma}$ in the heavy quark limit.
As the mass expansion (3), the square of $f$ can be expanded in the same way,
\begin{equation}
f^2=\displaystyle\bar{f}^2+\frac{{\delta f^K}^2}{m_Q}
+\frac{{\delta f^G}^2}{m_Q}<\vec{s}_Q\cdot\vec{j}_l>+O(\frac{1}{m_Q^2})~,
\end{equation}
where $\bar{f}^2$ denotes the leading order result and
${\delta f^K}^2$ and ${\delta f^G}^2$ the spin-conserved and spin-violated
$1/m_Q$ corrections respectively.\par
\vspace{1.0cm} 
   The two-point correlator $\Gamma(\omega)$ which we choose for sum rule
analyzing 
in the HQET is
\begin{equation}
\Gamma_{ij}(\omega)=i\int d^4xe^{ikx}<0|T\tilde{j}^v_i(x)\bar{\tilde{j}}^v_j
(0)|0>,~~~~i, j=1,2,
\end{equation}
where $\omega=2v\cdot k$.  The hadronic representation of this correlator is
\begin{equation}
\Gamma_{ij}(\omega)=
%\frac{2<0|\tilde{j}^v_i|{\cal B}>
%<{\cal B}|\bar{\tilde{j}}^v_j|0>}{2\tilde{\Lambda}-\omega}+{\rm res.}~,
(\frac{2\bar{f}^2}{2\bar{\Lambda}-\omega}
-\frac{1}{m_Q}\frac{4\bar{f}^2\delta\bar{\Lambda}}{(2\bar{\Lambda}-\omega)^2}
+\frac{1}{m_Q}\frac{2\delta f^2}{2\bar{\Lambda}-\omega})_{ij}
\frac{1+\not v}{2}+{\rm res.}~,
\end{equation}
%where ${\cal B}$ denotes the ground state of the heavy baryon. 
where $\delta\Lambda$ and $\delta f^2$ stand for the $1/m_Q$ corrections
in Eqs. (3) and (9).  
On the other hand, 
$\Gamma_{ij}(\omega)$ can be calculated in terms of quark and gluon language
with vacuum condensates.  This establishes the sum rule.  We use the
commonly adopted quark-hadron duality for the resonance part of Eq. 
(11), 
\begin{equation}
{\rm res.}=\frac{1}{\pi}\int_{\omega_c}^{\infty}d\omega'
\frac{{\rm Im}~\Gamma^{\rm pert}_{ij}(\omega')}{\omega'-\omega}~,
\end{equation}
where $\Gamma^{\rm pert}_{ij}(\omega)$ denotes the perturbative contribution,
and $\omega_c$ is the continuum threshold.
In this work, we shall consider only the diagonal correlators ($i=j$).\par
\vspace{1.0cm} 
   The calculations of $\Gamma(\omega)$ are straightforward.  The fixed point
gauge is used [11].  
All the condensates with dimensions lower than 6 are retained.
We also include the dimension 6 condensate $<\bar{q}(0)q(x)>^2$ in our
analysis which 
is a main contribution.  We use the gaussian ansatz for the distribution
in spacetime for this condensate [12].
In the heavy quark limit, we have double checked the analysis of Ref. [7].  
We use the following values of the condensates,
\begin{equation}
\begin{array}{rcl}
<\bar{q}q>&\simeq&-(0.23~ {\rm GeV})^3~,\\
<\alpha_sGG>&\simeq&0.04~ {\rm GeV}^4~,\\
<g\bar{q}\sigma_{\mu\nu}G^{\mu\nu}q>&\equiv&m_0^2<\bar{q}q>~,
~~~~~~m_0^2\simeq0.8~{\rm GeV}^2~.\\
\end{array}
\end{equation}
When $\omega_c$ lies between $2.1-2.7$ GeV for $\Lambda_Q$ and between
$2.3-2.9$ GeV for $\Sigma_Q^{(*)}$, the stability window exists.  
We obtain
\begin{equation}
\begin{array}{rcl}
\bar{\Lambda}_{\Lambda}&=&0.79\pm0.05~{\rm GeV}~,\\
\bar{f}^2_{\Lambda}&=&(0.3\pm0.1)\times 10^{-3}~{\rm GeV}^6~,
\end{array}
\end{equation}
for $\Lambda_Q$ baryon; and 
\begin{equation}
\begin{array}{rcl}
\bar{\Lambda}_{\Sigma}&=&0.96\pm0.05~{\rm GeV}~,\\
\bar{f}^2_{\Sigma}&=&(1.7\pm0.5)\times 10^{-3}~ {\rm GeV}^6~,
\end{array}
\end{equation}
for $\Sigma_Q^{(*)}$ baryon.  The normalization 
${\rm Tr} \tau^{\dagger}\tau=1$ has been used in the analysis.
The errors quoted in Eqs. (14) and (15) contain only those from the stability
of the sum rule windows.  We still do not know the $\alpha_s$ corrections
for baryons.  Taking the meson system as a reference [4], the $\alpha_s$
correction is very small for $\bar{\Lambda}$, however it is very large (30\%)
for $\bar{f}$.  
The numerical results are in agreement with that of Ref. [7], the range
of the Borel parameter is the same $T=0.4-0.7$.\par
\vspace{1.0cm}
   The $1/m_Q$ corrections to the two-point correlator $\Gamma(\omega)$ can be
calculated by including insertions of the $1/m_Q$ operators of the Lagrangian 
(2) with
standard method which is shown in Fig. 1.  The insertions
of spin-conserved and spin-violated operators are calculated separately.
The final
form of the sum rules are obtained by performing Borel transformation.  
With some simple tricks [13], the sum rules for the 
mass and $f$ can be
separated.
   The results for the mass of $\Lambda_Q$ baryon come from the 
spin-conserved operators only
($\delta\Lambda_{\Lambda}=\delta\Lambda_{\Lambda}^K$):
\begin{equation}
\begin{array}{rcl}
\delta\Lambda_{\Lambda_1}&=&\displaystyle
-\frac{T^2}{16\bar{f}^2_{\Lambda}}\frac{d}{dT}
I_{\Lambda_1}~,\\[4mm]
\delta\Lambda_{\Lambda_2}&=&\displaystyle
-\frac{T^2}{16\bar{f}^2_{\Lambda}}\frac{d}{dT}
I_{\Lambda_2}~,
\end{array}
\end{equation}
where
\begin{equation}
\begin{array}{rcl}
I_{\Lambda_1}&=&\displaystyle(\frac{3}{2^7\pi^45}\int_{0}^{\omega_c}d\omega
\omega^6 e^{-\omega/T}+\frac{m_0^2<\bar{q}q>^2}{T}e^{-\frac{m_0^2}{2T^2}}
+\frac{43<\alpha_sGG>}{2^5\pi^39}T^3)e^{2\bar{\Lambda}/T}~,\\[4mm]
I_{\Lambda_2}&=&\displaystyle(\frac{1}{2^7\pi^4}\int_{0}^{\omega_c}d\omega
\omega^6 e^{-\omega/T}+\frac{m_0^2<\bar{q}q>^2}{T}e^{-\frac{m_0^2}{2T^2}}
+\frac{61<\alpha_sGG>}{2^5\pi^39}T^3)e^{2\bar{\Lambda}/T}~,
\end{array}
\end{equation}
and the subscripts 1 and 2 denote $\tilde{j}^v_1$ and $\tilde{j}^v_2$
respectively.  The sum rule for $f$ of $\Lambda_Q$ is
\begin{equation}
\delta f^2_{\Lambda_i}=-\frac{1}{8}(1+\frac{d}{d\ln T})I_{\Lambda_i}~.
\end{equation}
\par
\vspace{1.0cm}
   The masses of baryons $\Sigma_Q$ and $\Sigma_Q^*$ 
are given in terms of $\delta\Lambda^K$ and $\delta\Lambda^G$.
They are determined by the following sum rules,
\begin{equation}
\begin{array}{rcl}
\delta\Lambda_{\Sigma_1}^K&=&\displaystyle
-\frac{T^2}{16\bar{f}^2_{\Sigma}}\frac{d}{dT}
I_{\Sigma_1}^K~,\\[4mm]
\delta\Lambda_{\Sigma_2}^K&=&\displaystyle
-\frac{T^2}{16\bar{f}^2_{\Sigma}}\frac{d}{dT}
I_{\Sigma_2}^K~,\\[4mm]
\delta\Lambda_{\Sigma_1}^G=\delta\Lambda_{\Sigma_2}^G&=&\displaystyle
-\frac{T^2}
{16\bar{f}^2_{\Sigma}}\frac{d}{dT}I_{\Sigma}^G~,
\end{array}
\end{equation}
where
\begin{equation}
\begin{array}{rcl}
I_{\Sigma_1}^K&=&\displaystyle(\frac{11}{2^7\pi^45}\int_{0}^{\omega_c}d\omega
\omega^6 e^{-\omega/T}+\frac{3m_0^2<\bar{q}q>^2}{T}e^{-\frac{m_0^2}{2T^2}}
+\frac{13<\alpha_sGG>}{2^5\pi^33}T^3)e^{2\bar{\Lambda}/T}~,\\[4mm]
I_{\Sigma_2}^K&=&\displaystyle(\frac{13}{2^7\pi^45}\int_{0}^{\omega_c}d\omega
\omega^6 e^{-\omega/T}+\frac{3m_0^2<\bar{q}q>^2}{T}e^{-\frac{m_0^2}{2T^2}}
-\frac{5<\alpha_sGG>}{2^5\pi^33}T^3)e^{2\bar{\Lambda}/T}~,\\[4mm]
I_{\Sigma}^G&=&\displaystyle
\frac{<\alpha_s GG>}{4\pi^3}T^3e^{2\bar{\Lambda}/T}~.
\end{array}
\end{equation}
The sum rules for $f$ are given by
\begin{equation}
\delta {f^2}_{\Sigma_i}^{K,G}=-\frac{1}{8}(1+\frac{d}{d\ln T})
I_{\Sigma_{(i)}}^{K,G}~.
\end{equation}
It can be seen that while the two
diagonal sum rules coincided with each other
at the leading order, they are no longer the same 
for the spin-conserved 
$1/m_Q$ corrections.\par
\vspace{1.0cm}
   The numerical sum rule results for the $1/m_Q$ corrections 
-- $\delta\Lambda$ and $\delta f^2$ in Eqs. (3) and (9) are given in 
Tables 1 and 2 and Figs. 2-4.  
The numerical differences resulting from the different choices of 
$\tilde{j}^v$
are not significant.  The values of $\omega_c$ are generally smaller than
the leading order results, but still lie in the allowed range of the
leading order results.  The lower limit of the Borel parameter 
$T=0.4$ is
determined by requiring that the condensates in Eqs. (17) and (20) have
less than 40\% contribution. 
The upper limit $T=0.6$ is obtained by requiring that the pole contribution 
is over 70\%.
This window is narrower than the leading order one.  
In the window $T=0.4-0.6$ the results for
$\delta\Lambda_{\Lambda}$ and $\delta\Lambda_{\Sigma}^K$ are comparatively
stable.  However from Fig. 4, we see that
$\delta\Lambda^G_{\Sigma}$ has no good stability in this window.
This is because we have not included the Feynman diagrams with internal 
gluon lines 
which are expected to be important for the spin-violated terms.
Therefore the value $\delta\Lambda^G_{\Sigma}$ in Table 1 is not 
reliable.  The errors quoted in Tables 1 and 2 again only refer to that 
from the stability of the sum rule windows.\par
\vspace{1.0cm}
   From $m_{\Lambda_c}$ and $m_{\Lambda_b}$ [14], we determine the heavy 
quark
masses $m_c=1.43\pm0.05$ GeV and $ m_b=4.83\pm0.07$ GeV.  These values 
give the following results,
\begin{equation}
m_{\Sigma_c}=2.52\pm0.08~{\rm GeV}~, m_{\Sigma_c^*}=2.55\pm0.08
~{\rm GeV}~,
\end{equation}
\begin{equation} 
m_{\Sigma_b}=5.83\pm0.09~{\rm GeV}~, m_{\Sigma_b^*}=5.84\pm0.09
~{\rm GeV}~.
\end{equation}
From the discussion above we know the individual mass value in Eqs. (22)
and (23) suffers from the inaccuracy of $\delta\Lambda^G_{\Sigma}$.  
The quantity 
\begin{equation}
\frac{1}{3}(m_{\Sigma_Q}+2m_{\Sigma_Q^*})=m_Q+\bar{\Lambda}
+\frac{1}{m_Q}(0.22\pm0.06~{\rm GeV}^2)
\end{equation}
is independent of $\delta\Lambda^G_{\Sigma}$, therefore more reliable.
It is $2.54\pm0.08$ GeV for $c$ quark case and $5.83\pm0.09$ GeV for $b$
quark case.  
Experimentally $m_{\Sigma_c}=2453\pm0.2$ MeV [14]. There is an
experimental evidence for $\Sigma_c^*$ at $m_{\Sigma_c^*}=2530\pm7$ MeV 
[14].  If we take this value for $m_{\Sigma_c^*}$, we have 
$\frac{1}{3}(m_{\Sigma_c}+2m_{\Sigma_c^*})=2504\pm5$ MeV.  This is 
in reasonable agreement with the theoretical value.
The corresponding quantity for the bottom quark can be checked by the experiments in the 
near future.\par
\vspace{1.0cm}
  To conclude, we have calculated the $1/m_Q$ corrections to the heavy
baryon
masses from the QCD sum rules within the framework of the HQET.  This study
refines the leading order analysis [7].  Furthermore within this framework,
we can study the three-point correlators which will give the form factors
for the weak transitions of the heavy baryons [8] to the order of $1/m_Q$.  
It is also viable to include the
QCD radiative corrections in the leading order and subleading order 
calculations.  Both of these two aspects are under our studying.\par

\vspace{2.0cm}

   One of us (Liu) would like to thank M. Chabab, K.T. Chao, 
W.F. Chen, Y. Liao, M. Tong and 
especially C.W. Luo for helpful discussions.
This work is supported in part by the China Postdoctoral Science 
Foundation.

\newpage

{\Large \bf Figure captions}\\

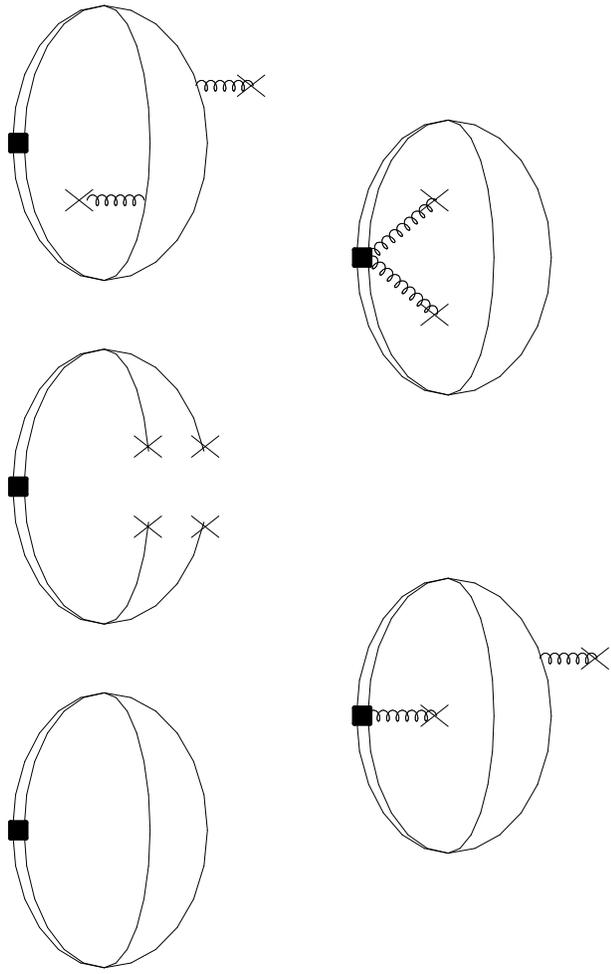
Fig. 1.  The subleading operator insertions relevant to our analysis.\\

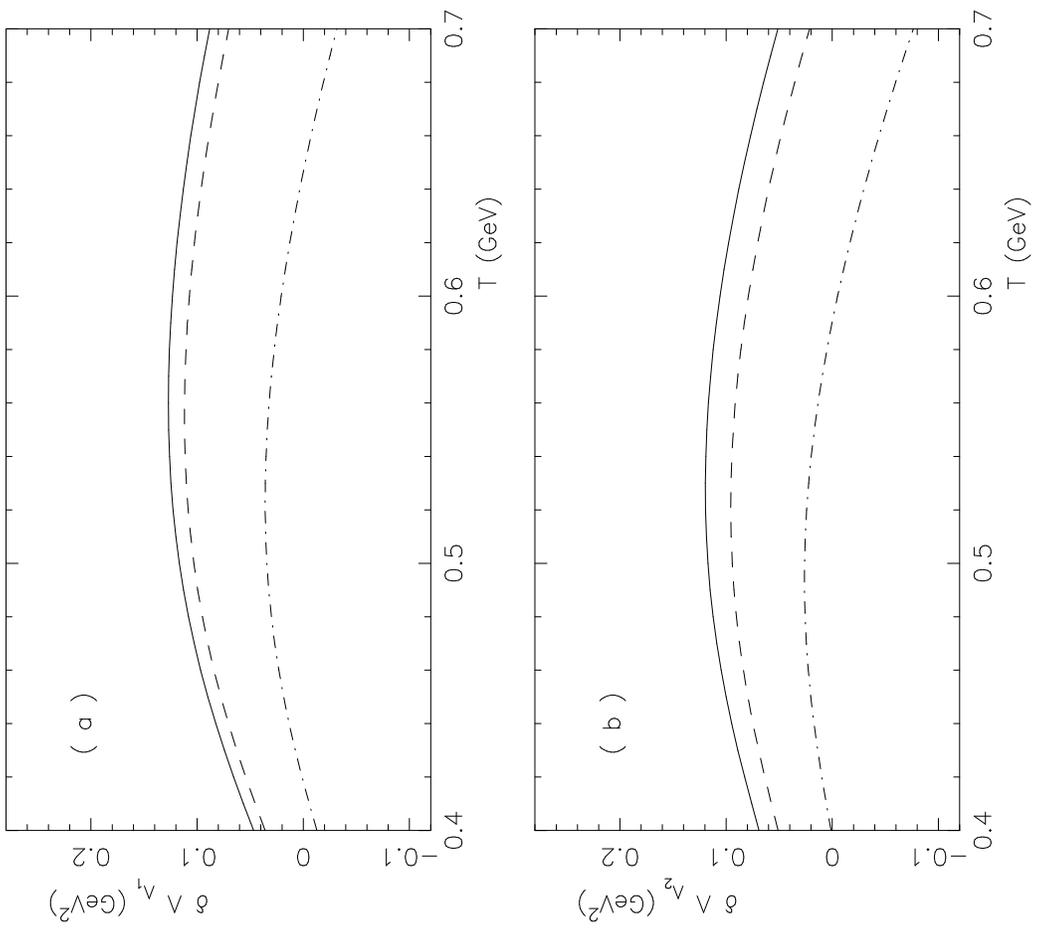
Fig. 2.  Sum rules for $\delta\Lambda_{\Lambda}$ with (a) $\tilde{j}^v_1$
and (b) $\tilde{j}^v_2$.  $\omega_c=2.0, 2.1, 2.4$ GeV for solid, dashed, 
dash-dotted curves respectively. The sum rule window is $T=0.4-0.6$ GeV.\\

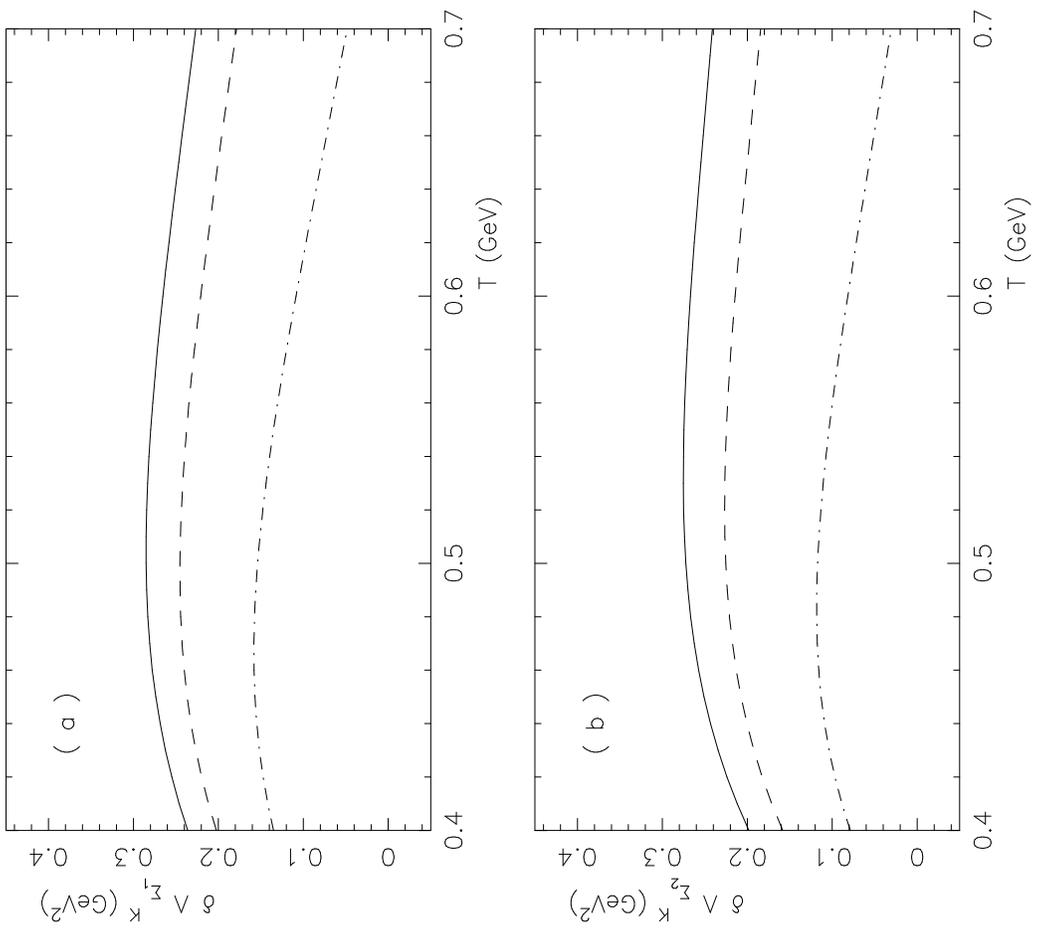
Fig. 3.  Sum rules for $\delta\Lambda_{\Sigma}^K$ with (a) $\tilde{j}^v_1$
and (b) $\tilde{j}^v_2$.  $\omega_c=2.2, 2.4, 2.7$ GeV for solid, dashed, 
dash-dotted curves respectively. The sum rule window is $T=0.4-0.6$ GeV.\\

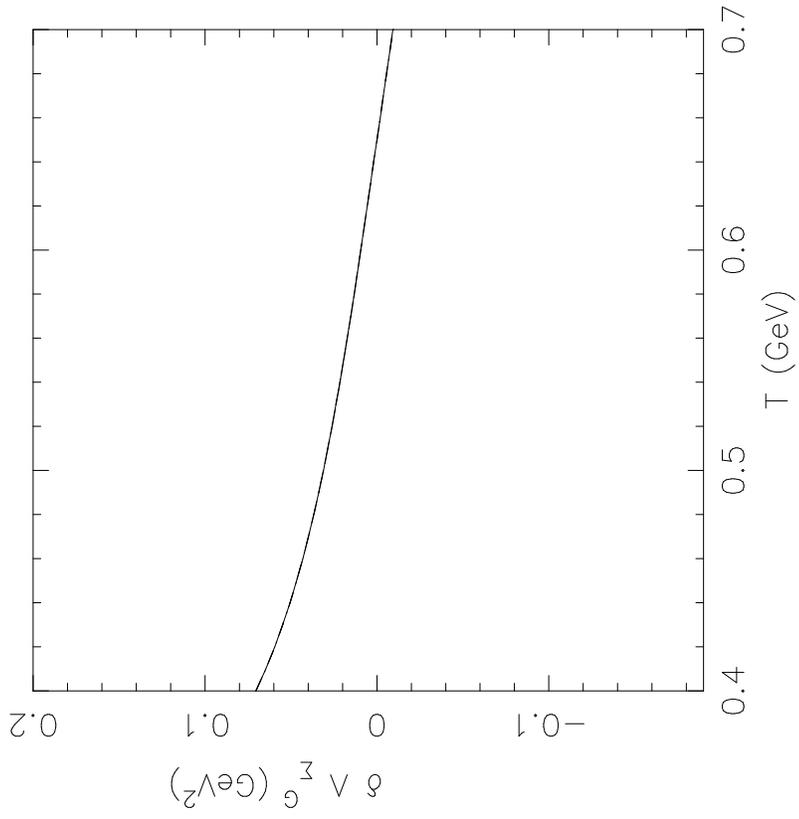
Fig. 4.  Sum rule for $\delta\Lambda_{\Sigma}^G$.
The sum rule window is $T=0.4-0.6$ GeV.

\newpage

{\Large \bf Tables}\\

Table 1.  Numerical results for $\delta\Lambda$.\\

\begin{tabular}{lccc}
\hline\hline 
 &$\delta\Lambda_{\Lambda}({\rm GeV}^2)$
&$\delta\Lambda_{\Sigma}^K({\rm GeV}^2)$
&$\delta\Lambda_{\Sigma}^G({\rm GeV}^2)$\\
 &$\omega_c$=$2.1\pm0.1$ GeV&$\omega_c$=$2.4\pm0.2$ GeV&\\
\hline
$\tilde{j}^v_1$&$0.09\pm0.03$&$0.22\pm0.06$&$0.03\pm0.02$\\
$\tilde{j}^v_2$&$0.09\pm0.05$&$0.21\pm0.06$&$0.03\pm0.02$\\
\hline\hline
\end{tabular}

\vspace{20mm}

Table 2.  Numerical results for $\delta f^2$.\\

\begin{tabular}{lccc}
\hline\hline 
 &$\delta f^2_{\Lambda}(10^{-3}{\rm GeV}^6)$
&${\delta f^K_{\Sigma}}^2 (10^{-3}{\rm GeV}^6)$
&${\delta f^G_{\Sigma}}^2 (10^{-3}{\rm GeV}^6)$\\
&$\omega_c$=$2.1\pm0.1$ GeV&$\omega_c$=$2.4\pm0.2$ GeV&\\
\hline
$\tilde{j}^v_1$&$-0.20\pm0.1$&$0.7\pm0.4$&$-0.1\pm0.1$\\
$\tilde{j}^v_2$&$-0.3\pm0.1$&$0.8\pm0.5$&$-0.1\pm0.1$\\
\hline\hline
\end{tabular}

\end{document}